\documentstyle[twoside,fleqn,espcrc2,epsf]{article}

\newcommand{\bee}{\begin{equation}}
\newcommand{\ee}{\end{equation}}
\newcommand{\beea}{\begin{eqnarray}}
\newcommand{\eea}{\end{eqnarray}}

\newcommand{\AmS}{{\protect\the\textfont2
  A\kern-.1667em\lower.5ex\hbox{M}\kern-.125emS}}

\title{Scaling Tests of Some Lattice Fermion Actions}

\author{T. DeGrand\address{Physics Department, 
        University of Colorado, \\ 
        Boulder, CO 80309 USA}}
       
\begin{document}

\begin{abstract}
I describe  studies of quenched spectroscopy, using crude approximations
to fixed point actions for fermions interacting with SU(3) gauge fields.
These actions have a hypercubic kinetic term and a complicated
lattice anomalous magnetic moment term.
They  show improved scaling compared to the conventional Wilson action.
\end{abstract}

\maketitle

\section{INTRODUCTION}

Fixed point actions of asymptotically free spin and gauge models
show improved scaling behavior compared to
standard discretizations.  A natural extension of this work is
to models with fermions interacting with gauge fields.
I have been doing nonperturbative studies of  fermion actions ``inspired''
by a perturbative fixed point (FP)
vertex\cite{FP}. I have been testing
scaling using quenched hadron spectroscopy.  The calculation is
not complete, but the results are encouraging.

The project is divided into two parts. First, one must construct
 approximate nonperturbative
FP actions valid for rough gauge configurations, and second,
one must perform scaling tests.
 I will describe the work in reverse order, however,
since one does not
 have to know much about FP actions to look at the scaling tests.

In order not to entangle
scaling tests with extrapolations to infinite volume or to
zero quark mass, all tests are done in
volumes of fixed physical size (defined through the critical temperature
for deconfinement--all my tests are on lattice size $L=2/T_c$) and at fixed physical quark mass (defined either
by interpolating lattice data to a fixed value of $m_\pi/m_\rho$
or to a fixed value of $m_\pi/T_c$).  The tests are done at very coarse
lattice spacing, where scaling violations from conventional actions
 are very large. Thus only modest statistics are required to identify
improvement.

Another scaling test is the meson or baryon dispersion relation.
In lattices of fixed physical volume set by $T_c$,
 the physical momenta corresponding
to the different allowed lattice modes are multiples of $T_c$,
$a\vec p = 2\pi \vec n/L$ or $\vec p = \pi T_c \vec n$, if $L=2/T_c$,
and one can compare data  at the same physical momentum
with different lattice spacings.
Wilson fermions at $\beta<6.0$ (using the Wilson gauge action)
exhibit bad scaling or rotational invariance violations,
which is often parameterized as ``$c^2_{eff} <1$,''
with $c^2_{eff}= (E(p)^2-m^2)/p^2 $.

\section{GENERIC ACTIONS}
All the actions I have tested are Wilson-like: they have four-component
spinors and no protection from additive mass renormalization.
The ``footprint'' of the actions covers a hypercube.
The free field limit of the action is parameterized as
\bee
\Delta_0(x) = \lambda(x) + \sum_\mu \gamma_\mu \rho_\mu(x)
\ee
with five nonzero $\lambda$'s and four nonzero $\rho$'s, corresponding
to each of the generic types of
offsets.  These parameters are set by finding
a particular FP action and truncating it to a hypercube.
They vary smoothly with the bare mass $m_0$ and in practice are parameterized
as  linear functions of $m_0$.

The actions are made gauge invariant by connecting the fermion fields on
different sites by sums of products of link variables. The actions differ in the
specific choice of composition of the gauge connections.

The actions  also include an anomalous magnetic moment term, or
``Pauli term,'' which is
a very complicated extension of the standard ``clover'' term.
Its effect is to correct lattice artifacts in magnetic interactions.
This term is part of the FP vertex.
It consists of a large number of contributions of the form
$\sum c(r) \bar \psi(x) \sigma_{\mu\nu}F_{\mu\nu} \psi(x+r)$.
The lattice analog of $F_{\mu\nu}$ is made of oriented paths connecting
sites $x$ and $x+r$.
We can parameterize the magnetic interactions 
by comparing the inverse of the magnetic moment, labeled
by the ``magnetic mass'' $m_B$, to the pole mass $m_0$.
My truncation of the FP vertex keeps  only terms in which the
fermion and antifermion lie in a cube, connected by minimum length gauge paths,
and normalized so that $m_0=m_B$.

The cost of any of the actions I am testing is about
 56 times as expensive as the usual Wilson action.
About half of that (20) is due to the kinetic term and the rest is due to
 the complicated Pauli term. All the  gauge connections are pre-computed.

\section{TESTS OF ACTIONS}

I have  tested several related actions.
All but one action use the ``instanton-friendly'' SU(3) gauge action of our
recent work\cite{INST}.
We have made rough measurements of its $\beta_c(N_t)$ for deconfinement
to set the scale.
All the actions use gauge connections which are basically averages
over the shortest paths connecting terms in the action.

The first action uses ``tadpole-improved fat links''
(action T).
I begin with a vertex made of 
scalar and vector terms using the shortest gauge paths
connecting the fermions, and then  replace the individual link
variable by a ``fat link,'' 
\beea
V_\mu(x) = (1-\alpha)U_\mu(x) + \nonumber \\
 {\alpha \over 6} \sum_\nu (
U_\nu(x)U_\mu(x+\hat \nu) U_\nu^\dagger(x+\hat\mu)  \nonumber \\
+ \dots),
\eea
 taking $\alpha=1/2$. The links are then tadpole-improved.

The second action  (action A) replaces all the links by multi-level APE-blocked links: iterate Eqn. 2,
with $V^0_j(x)=U_j(x)$ and $V^{n+1}_j(x)$ projected back onto $SU(3)$.
I used $\alpha=0.3$ 
and seven blocking steps.

A third action (action F)
uses different levels of APE-blocking for the links
participating in the scalar and vector  (``kinetic'') part of the action
compared to the links used in the Pauli terms.
The kinetic links have one level of APE-blocking with $\alpha=0.5$.
The links making up the
Pauli terms are replaced by ten-level APE blocked links.

\begin{figure}[htb]
\begin{center}
\vskip -10mm
\leavevmode
\epsfxsize=60mm
\epsfbox[40 50 530 590]{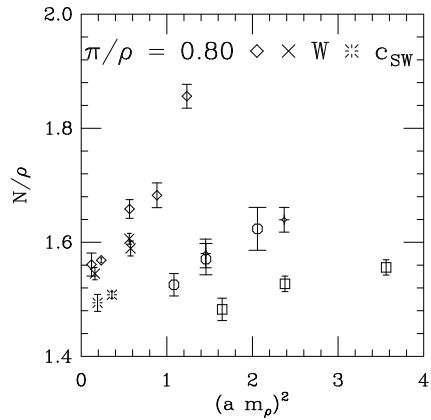}
\vskip -5mm
\end{center}
\caption{A scaling test for new actions: (squares action F,
 octagons action A,  fancy diamond action T)
vs. Wilson actions on lattices of fixed physical size (diamonds)
and larger volumes (crosses). Data are interpolated to
$\pi/\rho=0.80$.}
\label{fig:ratio80}
\end{figure}

Figure \ref{fig:ratio80} shows a comparison of
Wilson action data and our test  actions for the $N/\rho$ mass ratio
at one fixed $\pi/\rho$ mass ratio. 
In this figure the diamonds are Wilson action data in lattices of fixed
physical size
 ($4^3$ at $\beta=5.1$,
$6^3$ at $\beta=5.54$,
$8^3$ at $\beta=5.7$,
$16^3$ at $\beta=6.0$
$24^3$ at $\beta=6.3$) and the crosses are data in various larger lattices.
The bursts are from the nonperturbatively improved Wilson action of Ref.
\cite{ALPHA}.
The other plotting symbols show our test actions.

The ``fat links'' considerably reduce the renormalization of bare parameters.
For example, the bare mass at which $m_\pi=0$ is $m_0=-0.4$
and -0.13 at $\beta_c(N_t=2)$ and $\beta_c(N_t=4)$ for action A,
in contrast to -1.58 and -1.04 for the standard Wilson action
($m_0= 1/2\kappa_c -4$).

\begin{figure}[htb]
\begin{center}
\vskip -10mm
\leavevmode
\epsfxsize=60mm
\epsfbox[40 50 530 590]{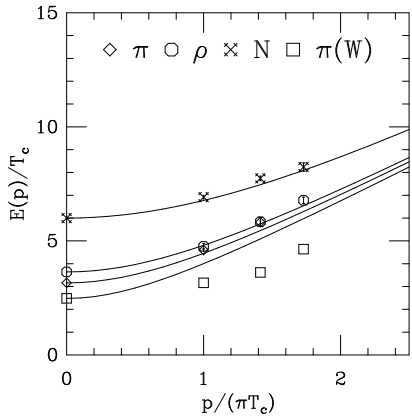}
\vskip -5mm
\end{center}
\caption{Dispersion relation for heavy hadrons at $aT_c=1/2$
 from a test action.
The curves are the continuum dispersion relation for the appropriate
(measured) hadron mass. A Wilson pseudoscalar is shown by the squares.}
\label{fig:eptc}
\end{figure}

I show one figure comparing dispersion relations.
The result for  test action T  is compared to the
Wilson action dispersion relation (for heavy pseudoscalars)
at $aT_c=1/2$ in Fig. \ref{fig:eptc}.
All of the test actions I have studied have good dispersion relations
even at $aT_c=1/2$. I believe that is a generic feature of the
hypercubic kinetic term.  Leaving out the Pauli term gives noticeable scaling violations
with a too-large $N/\rho$ ratio; probably one needs to keep some kind of
explicit Pauli/clover term in the lattice action
to boost the hyperfine splittings.

It appears that these actions are members of  a  family of actions which show
improved scaling, even at $\beta_c(N_t=2)$, about 0.4 fm lattice spacing.
My data by themselves do not suggest a unique way to extrapolate to
$a=0$.
However, if I linearly
extrapolate the $\pi/\rho=0.8$ Wilson    results to $a=0$,
I get $N/\rho$=1.44(2),
to be contrasted with quadratic extrapolations of
1.48(3) for the improved Wilson \cite{ALPHA},
1.42(6) for action A, and
and 1.44(3) for action F.

\section{TOWARDS A FP ACTION}
These actions come from a program to find a nonperturbative 
FP action\cite{FP}.
The present parameterizations began with a particular choice for an RGT.
We then constructed (semi-analytically) a FP vertex, or FP action valid
for smooth gauge configurations.
By solving the RG equations, the vertex can then be used to find approximate
actions valid for rough gauge configurations.
The FP vertices
we have found are far too complicated to  simulate, and so I must
truncate them before doing simulations.
 The fit to a FP action gradually worsens
as I move towards the stronger couplings of interest to simulations.
 I believe
a more sophisticated construction of the gauge connections is needed.
This may involve looking at the coupling of instantons and fermions.
This problem is under study.

\section{TENTATIVE CONCLUSIONS}
My goal is an action which scales at  $aT_c = 1/2$.
I clearly don't have that (yet). However, it appears that I have gained a factor
of perhaps 3 in lattice spacing for an equivalent amount of scale violation,
compared to the Wilson action.

The generic feature of these actions which might be interesting to other
``improvers'' is the use of some sort of fat link in the gauge
connections. The nonlocal Pauli terms are needed to satisfy the FP
equations, but it is not clear to me that replacing the
complicated Pauli term by an appropriately
normalized simple clover term will seriously harm scaling.

This work was supported by the U.S. Department of 
Energy, with some computations done on the T3E at Pittsburgh Supercomputing
Center through resources awarded to the MILC collaboration.

\end{document}